\documentclass{article}

\usepackage{arxiv}

\usepackage[utf8]{inputenc}
\usepackage[T1]{fontenc}    
\usepackage{hyperref}       
\usepackage{url}            
\usepackage{booktabs}       
\usepackage{amsfonts}       
\usepackage{amsmath}
\usepackage{nicefrac}       
\usepackage{microtype}                   
\usepackage{graphicx}
\usepackage{verbatim}

\usepackage{amsmath,amssymb,amsthm}

\newtheorem{definition}{Definition}[section]

\usepackage{tikz}
\usetikzlibrary{calc}
\usetikzlibrary{shapes.geometric, positioning, arrows.meta}

\usepackage{multirow}

\usepackage{enumitem}
\setlist[itemize]{labelindent=1em,leftmargin=2em,labelsep=0.5em}
\usepackage{braket}
\usepackage[english]{babel}
\usepackage{microtype}

\usepackage{array}
\usepackage{tabularx}

\newcolumntype{L}[1]{>{\raggedright\arraybackslash}p{#1}}
\newcolumntype{Y}{>{\raggedright\arraybackslash}X}

\makeatletter
\def\blfootnote{\xdef\@thefnmark{}\@footnotetext}
\makeatother

\usepackage{fancyhdr}
\begin{document}

\title{Beyond the Canonical Protocol: Quantum Encrypted Cloning from Secret-Sharing Access Structures}

\author{
  \href{https://orcid.org/0000-0001-5186-0199}{\includegraphics[scale=0.06]{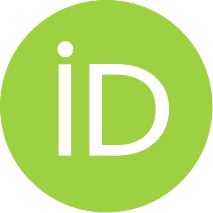}\hspace{1mm} Gabriele Gianini}\\
%  Universit\`a degli Studi di Milano-Bicocca \\
  University of Milano-Bicocca \\
  Milan, Italy \\
  \texttt{gabriele.gianini@unimib.it}\\ \vspace{1pt}
\and
  \href{https://orcid.org/0000-0003-1737-6218}{\includegraphics[scale=0.06]{orcid.pdf}\hspace{1mm}
  Stelvio Cimato}\\
%  Universit\`a degli Studi di Milano\\
  University of Milano\\
  Milan, Italy\\
  \texttt{stelvio.cimato@unimi.it}\\ \vspace{1pt}
\and
  \href{https://orcid.org/0000-0002-3299-448X}{\includegraphics[scale=0.06]{orcid.pdf}\hspace{1mm}
  Jianyi Lin}\\
%  Universit\`a Cattolica del Sacro Cuore\\
  Universit\`a Cattolica\\
  Milan, Italy\\
  \texttt{jianyi.lin@unicatt.it}\\ \vspace{1pt}
\and
\href{https://orcid.org/0000-0002-9585-7810}{\includegraphics[scale=0.06]{orcid.pdf}\hspace{1mm}
  Omar Hasan}\\
%  Institut National des Sciences Appliqu\'ees de Lyon\\
  INSA of Lyon\\
  Lyon, France \\
  \texttt{omar.hasan@liris.cnrs.fr}\\ \vspace{1pt}
\and
  \href{https://orcid.org/0000-0002-1087-4866}{\includegraphics[scale=0.06]{orcid.pdf}\hspace{1mm}
  Corrado Mio}\\
%  Khalifa University of Science and Technology\\
  Khalifa University\\
  Abu Dhabi, UAE\\
\texttt{corrado.mio@ku.ac.ae}\\ \vspace{1pt}
\and
  \href{https://orcid.org/0000-0002-9557-6496}{\includegraphics[scale=0.06]{orcid.pdf}\hspace{1mm}
  Ernesto Damiani}\\
  University of Milano\\
  Milan, Italy\\
  \texttt{ernesto.damiani@unimi.it}
}%%%%%%%%%%%% END AUTHOR
\date{}
\maketitle
%%%%%%%%%%%%%%%%%%%%%%%%%%%%%%%%%%%%%%%%%%%%%%%%%%%%%%%%%%%
\begin{abstract}
Quantum encrypted cloning shows that an unknown quantum state can be distributed into multiple encrypted copies without contradicting the no-cloning theorem: each copy is unusable on its own, but can be redeemed together with a suitable quantum key. Recent work has related canonical encrypted-cloning protocols to particular forms of quantum secret sharing. Here we take the converse perspective: instead of mapping a given encrypted-cloning protocol into QSS, we use QSS access structures as a design library from which encrypted-cloning schemes can be extracted.
The criterion is access-structural. A QSS scheme supports a quantum encrypted-cloning structure whenever it contains a family of qualified sets with a non-qualified common intersection. The common subsystem is interpreted as the key, while the non-common parts are interpreted as encrypted clones relative to that key. Thus quantum encrypted cloning does not require a new notion of recoverability beyond QSS; what changes is the operational reading of QSS constituents as a mechanism for delayed and alternative redemption opportunities.
This viewpoint separates redemption from perfect secrecy. Perfect QSS yields encrypted-cloning schemes with forbidden non-qualified subsystems, whereas ramp QSS naturally allows intermediate, partially informative non-redeeming subsystems. The resulting framework broadens quantum encrypted cloning from a specific protocol to a general access-structure primitive. We illustrate the extraction principle with threshold-like, ramp, hierarchical, and compartmented architectures, showing how encrypted clones may be symmetric or asymmetric, individual or composite, perfectly hidden or leaky. Equivalently, these constructions can be viewed as overlapping erasure-recovery regions of an isometric quantum code. This establishes secret sharing as a systematic design language for encrypted quantum redundancy.

\end{abstract}
%%%%%%%%%%%%%%%%%%%%%%%%%%%%%%%%%%%%%%%%%%%%%%%%%%%%%%%%

\section{Introduction}\label{sec:intro}

The no-cloning theorem is one of the most distinctive constraints of quantum information theory: an unknown quantum state cannot be copied perfectly by any physical process. This impossibility rules out the direct duplication of quantum information, but it does not rule out all forms of quantum redundancy. A recent example is provided by Quantum Encrypted Cloning (hereafter, QECL), recently introduced by Yamaguchi and Kempf~\cite{yamaguchi2026encrypted} (and extended to qudits by Cear\u{a}~\cite{ceara2026cloningencryptedquantumstates}).
In this primitive, an unknown quantum state is distributed into several encrypted quantum systems. None of these systems is a usable copy on its own, but each of them can be redeemed into the original state when combined with a suitable quantum key. The different redemptions are therefore alternative rather than independent, which keeps the construction compatible with the no-cloning theorem.

Yamaguchi and Kempf already pointed out that QECL is closely related to Quantum Secret Sharing (QSS)~\cite{hillery1999quantum,cleve1999share,gottesman2000theory}. Indeed, the pair consisting of the key and one encrypted clone behaves like an authorized set: together they can reconstruct the original secret. This observation suggests that encrypted cloning should be understood also as an access-structure construction. Recent work has mapped specific instances of the canonical encrypted-cloning protocols and their higher-dimensional extensions to threshold QSS \cite{lim2026encrypted} (details reported in Section \ref{sec:discussion}). 

The perspective developed here is complementary. Rather than starting from a given encrypted-cloning protocol and mapping it into QSS, we start from QSS access structures and provide a criterion to generate from them a QECL interpretation.
Our contribution is therefore an operational reinterpretation of selected QSS access structures: qualified sets with a common intersection are viewed as alternative key--clone redemption sets, where the intersection plays the role of the key and the remainder of the qualified set plays the role of the clone. 

However, the relation with QSS is subtle. In perfect QSS, unauthorized sets are required to be completely ignorant of the secret. By contrast, the operational requirement of encrypted cloning is weaker: the essential condition is that each key--clone pair can redeem the state, not necessarily that every non-redeeming subsystem is completely blind.
This distinction is important. In recent work~\cite{gianini2026encrypted,gianini2026full}, we showed that certain instances of the Yamaguchi--Kempf construction exhibit patial leakage. Such leakage rules out a direct interpretation as a perfect threshold QSS scheme, but it does not invalidate the encrypted-cloning functionality, although it changes the security interpretation of the scheme.
From the point of view of encrypted cloning, what matters is redundant single redemption: several subsystems can each complete the key into a recovery set, while no collection of disjoint systems can be used to obtain independent clear copies of the unknown state. Thus, leakage is fatal to a perfect-QSS interpretation, but not necessarily to encrypted cloning.

This motivates the access-structure viewpoint developed in this work. We separate the \emph{redemption property} from the \emph{perfect-secrecy property}. The redemption property says that certain subsets can reconstruct the secret. The perfect-secrecy property says that all non-authorized subsets are completely independent of the secret. Perfect QSS imposes both conditions in a dichotomic way: subsets are either qualified or forbidden. Ramp QSS relaxes this dichotomy by allowing intermediate subsets that may contain partial information without being able to reconstruct the secret~\cite{ogawa2005quantum,zhang2015strongly,matsumoto2014coding}. This is precisely the kind of structure needed to describe encrypted cloning schemes in which non-redeeming subsystems may leak information without becoming redeeming subsystems.

The main claim of this paper is that \textit{QECL can be formulated as a key-centered substructure of a QSS access structure}.

This perspective shifts the role of QSS. Rather than being only an analogy for encrypted cloning, QSS becomes a design language for constructing encrypted-cloning schemes. For instance:\ threshold schemes give rise to symmetric encrypted clones by choosing a key of one share less than the reconstruction threshold; ramp schemes naturally produce leaky encrypted-cloning structures, where intermediate subsets may retain partial information; hierarchical, compartmented, graph-state, and stabilizer-based QSS schemes suggest asymmetric and multi-key variants, in which different classes of encrypted clones complete different keys. In this sense, \textit{QSS access structures provide an architectural library for encrypted quantum cloning protocols}.

It is worth mentioning that the same framework also admits a quantum error-correction interpretation~\cite{knill1997theory,nielsen2010quantum}. For an isometric encoding, recovery from an authorized set is equivalent to correction of the erasure of its complement. Thus, a family of key--clone redemption sets may equivalently be viewed as a family of overlapping erasure-recovery regions. 

The contributions of this work are as follows:
\begin{itemize}
    \item We extract from the canonical Yamaguchi--Kempf protocol a structural notion of QECL based on key--clone redemption sets, rather than on a specific circuit construction.
    
    \item We clarify the relation between encrypted cloning and QSS by separating \textit{redemption} from \textit{perfect secrecy}. This distinction explains why leakage in non-redeeming subsystems may rule out perfect threshold QSS while remaining compatible with encrypted cloning.
    
    \item We formulate QECL as a key-centered access substructure of a QSS scheme: a family of qualified sets with a common intersection interpreted as the key, and non-common parts interpreted as encrypted clones.
    
    \item We show how different classes of QSS architectures, including threshold, ramp, hierarchical and compartmented can be exported into encrypted-cloning architectures.
    
    \item We illustrate the framework through minimal few-qubit examples, showing how familiar access structures can be reinterpreted as QECL schemes.
\end{itemize}

The remainder of the paper is organized as follows. Section~\ref{sec:canonical-qecl} reviews the canonical QECL protocol and identifies the operational features that will be abstracted in the rest of the work. Section~\ref{sec:qecl-structures} extracts from this protocol a general structural notion of encrypted cloning based on key--clone redemption sets with a common intersection. Section~\ref{sec:qss-access} recalls the access-structure language of QSS, including perfect and ramp schemes and shows how encrypted cloning can be understood as a key-centered access substructure. Section~\ref{sec:exporting-qss} discusses how standard QSS architectures can be exported to encrypted cloning by selecting suitable overlapping families of qualified sets and presents minimal few-qubit examples illustrating the framework. Finally, Section~\ref{sec:discussion} discusses the contributions and outlines possible extensions.

%%%%%%%%%%%%%%%%%%%%%%%
% SECTION\ 2

\section{The canonical QECL protocol}
\label{sec:canonical-qecl}

We begin by recalling the Pauli-based encrypted-cloning protocol introduced by Yamaguchi and Kempf~\cite{yamaguchi2026encrypted}. The purpose of this section is to isolate the encoding and decoding structure that will later be reinterpreted in access-structure terms.

Let \(A\) denote the input qubit, prepared in an unknown pure state
\(
    \ket{\psi_0}_{A}
\), thus with density $(\rho_0)_{A}=\big(\ket{\psi_0}\bra{\psi_0}\big)_{A}$.
To generate \(n>1\) encrypted clones, the protocol introduces \(n\) Bell pairs
\begin{equation}
\ket{\phi}_{S_iK_i}
    =
    \frac{1}{\sqrt{2}}
    \left(
        \ket{00}+\ket{11}
    \right),
    \qquad i=1,\ldots,n.
\end{equation}
Here \(S_i\) is the \(i\)-th signal qubit, which will play the role of an encrypted clone, while \(K_i\) is the corresponding key qubit. The initial state is therefore
\begin{equation}
\ket{\psi_0}_{A}
    \otimes
    \bigotimes_{i=1}^n
    \ket{\phi}_{S_iK_i}.
\end{equation}
The encoding acts jointly on the input qubit \(A\) and on the signal qubits \(S_1,\ldots,S_n\), while leaving the key qubits \(K_1,\ldots,K_n\) untouched. It is defined by the unitary

\begin{equation}
U_{\mathrm{enc}}^{(n)}
    =
    \frac{1}{2}
    \sum_{\mu=0}^3
    \alpha_\mu^{-1}
    \sigma_\mu^{(A)}
    \otimes
    \left(
        \bigotimes_{i=1}^n
        \sigma_\mu^{(S_i)}
    \right),
\end{equation}

where
$
\sigma_0=I,\;
    \sigma_1=X,
    \;
    \sigma_2=Y,
    \;
    \sigma_3=Z
$,
and
$
\alpha_0=1,
    \;
    \alpha_1=\alpha_3=i,
    \;
    \alpha_2=-i^{\,n+1}
$.
The encoded state is
\begin{equation}
\ket{\Psi_{\mathrm{enc}}^{(n)}}
    =
    \frac{1}{2}
    \sum_{\mu=0}^3
    \alpha_\mu^{-1}
    \sigma_\mu^{(A)}\ket{\psi}_A
    \otimes
    \left(
        \bigotimes_{i=1}^n
        \ket{\phi_\mu}_{S_iK_i}
    \right),
\end{equation}
where 
\begin{equation}
\ket{\phi_\mu}_{S_iK_i}
    =
    \left(
        \sigma_\mu^{(S_i)}
        \otimes
        I^{(K_i)}
    \right)
    \ket{\phi}_{S_iK_i}.
\end{equation}
Thus the encoding creates a coherent superposition of four Pauli branches. In each branch, the same Pauli label \(\mu\) is coherently correlated with the transformed input qubit \(A\) and with all signal--key Bell pairs. The signal qubits \(S_i\) are therefore encrypted systems whose usefulness depends on access to the appropriate key degrees of freedom.

The corresponding encoded density matrix is
\begin{equation}
\rho_{\mathrm{enc}}^{(n)}
    =
    \frac{1}{4}
    \sum_{\mu,\nu=0}^3
    \alpha_\mu^{-1}\alpha_\nu
    \left(
        \sigma_\mu^{(A)}
        \ket{\psi}\!\bra{\psi}
        \sigma_\nu^{(A)}
    \right)
    \otimes
    \left(
        \bigotimes_{i=1}^n
        \ket{\phi_\mu}\!\bra{\phi_\nu}_{S_iK_i}
    \right).
\end{equation}

This expression will be useful later when discussing reduced states, but the core structural point is already visible at the level of the pure encoded state: the information about \(\ket{\psi}\) is distributed coherently across the input qubit $A$, the signal qubits $S_i$, and the key qubits $K_i$.

The decoding stage selects one signal qubit \(S_i\) and combines it with a suitable collection of key qubits. Operationally, the key information identifies and reverses the Pauli branch that encrypted the selected signal. Thus, the pair consisting of \(S_i\) and the appropriate key subsystem acts as a redemption set: it can recover the original unknown state.

The signal qubits are not independent copies; they are \textit{alternative encrypted access points} to the same quantum information.
This is the operational pattern that will be abstracted in the following sections. In the canonical protocol, the redemption sets have a specific signal--key structure. We will instead ask which features of this structure are essential. The answer proposed here is that \textit{encrypted cloning arises whenever a family of authorized sets has a nontrivial common intersection, interpreted as the key, while the non-common parts are interpreted as encrypted clones relative to that key}.

%%%%%%%%%%%%%%%%%%%%%%%%%%%%
% SECTION\ 3
%

\section{From the canonical protocol to QECL structures}
\label{sec:qecl-structures}

The canonical protocol reviewed in the previous section has a specific signal--key architecture. It contains signal qubits \(S_i\), key qubits \(K_i\), and authorized decoding sets determined by the signal--key pairing. However, not all details of that construction are essential for the encrypted-cloning interpretation. What matters operationally is that there are \textit{several alternative ways of recovering the same unknown input state}, and that each recovery requires combining a clone-like subsystem with a suitable key subsystem. 
This motivates the following abstraction. Let
\begin{equation}
V:\mathcal H_{\mathrm{in}}
    \longrightarrow
    \bigotimes_{j=1}^N \mathcal H_j
\end{equation}
be an isometric encoding of the input quantum system into \(N\) output subsystems. For any subset \(X\subseteq [N]\), the reduced channel available to \(X\) is
\begin{equation}
    \mathcal E_X(\rho_0)
    =
    \operatorname{Tr}_{\overline X}
    \left(
        V\rho_0 V^\dagger
    \right),
\end{equation}
where \(\overline X=[N]\setminus X\). We say that \(X\) is \emph{qualified} if there exists a decoding channel \(\mathcal D_X\), acting only on the subsystem \(X\), such that
\begin{equation}
    \mathcal D_X
    \left[
        \mathcal E_X(\rho_0)
    \right]
    =
    \rho_0
\end{equation}
for every input state \(\rho_0\). Thus a qualified set is a subsystem from which the encoded input can be perfectly recovered.

A QECL structure is obtained when the isometry admits a family of qualified sets
\(Q_1,\ldots,Q_m\)
with a nontrivial common intersection
\(K = \bigcap_{i=1}^m Q_i .\)
The subsystem \(K\) is interpreted as the quantum key. For each qualified set \(Q_i\), we define the corresponding non-common part by
\(S_i = Q_i\setminus K .\)
Then
\(Q_i = K\cup S_i ,\)
and \(S_i\) is interpreted as an encrypted clone relative to the key \(K\). The pair \(K\cup S_i\) is a redemption set: it is able to recover the original input state.

\begin{definition}[QECL access substructure]
Let \(\mathcal F\subseteq \mathcal P(\Omega)\) be the qualified-set family of an isometric QSS encoding. A QECL substructure is a pair \((K,\{S_i\}_{i=1}^m)\), with \(m\ge 2\), such that
\[
K\notin\mathcal F,\qquad S_i\notin\mathcal F,\qquad K\cup S_i\in\mathcal F
\]
for all \(i\), and
\[
K=\bigcap_i (K\cup S_i).
\]
The subsystem \(K\) is the key, and the \(S_i\)'s are encrypted clones relative to \(K\).
\end{definition}

This definition captures the structural content of the Yamaguchi--Kempf protocol. In that case, the encrypted clones are the signal qubits \(S_i\), while the key subsystem is built from the key qubits \(K_i\). Each signal qubit becomes useful only when combined with the appropriate key information. The signal qubits \textit{are therefore not copies in the ordinary sense}: they are alternative access points to the same encoded quantum information.

Notice that in the canonical protocol, the subsystem interpreted as the key may be a composite block rather than a single physical qubit. The present abstraction does not require the key to be elementary; it requires only that the selected redemption sets share a common non-qualified subsystem. Thus, when applying the criterion to the canonical construction, one must first specify the family of redemption sets under consideration and the corresponding common key block.

The requirement that the qualified sets overlap on \(K\) is crucial. If two qualified sets were disjoint, each of them could independently recover the same unknown quantum state, contradicting the no-cloning theorem. In a QECL structure, by contrast, the recovery sets are not independent: they all share the key subsystem. 

Thus an encrypted clone is a non common part of a qualified set, relative to a common key \(
    K=\bigcap_i Q_i 
\).
Thus the clone \(S_i\) completes the key \(K\) into a qualified set.
To exclude trivial cases it is required that the key alone is not qualified:
\(K \notin {\mathcal F} ,\)
where \({\mathcal F}\) denotes the family of qualified sets. Similarly, one may require that the encrypted clones alone are not qualified:
\(S_i \notin {\mathcal F}\).

Combinatorially, a QECL structure is a star-like pattern in the Boolean lattice of subsystems. Let \(\Omega\) be the set of  registers and let \({\mathcal F}\subseteq\mathcal P(\Omega)\) be the monotone family of qualified sets. A QECL substructure consists of a non-qualified subsystem \(K\) and a family of non-qualified subsystems \(S_1,\ldots,S_m\) such that
$
K\notin{\mathcal F},
    \;
    S_i\notin{\mathcal F},
    \;
    K\cup S_i\in{\mathcal F}
$
for every \(i\). The sets \(Q_i=K\cup S_i\) are the redemption sets. The subsystem \(K\) is interpreted as the key, while the \(S_i\)'s are interpreted as encrypted clones. Thus each \(S_i\) completes the key into a qualified set, although neither the key nor the clone is qualified on its own.

The status of the non-qualified sets is deliberately left open at this stage. In a perfectly secret realization, non-qualified subsystems may be completely independent of the input state. In a leaky realization, they may retain partial information without being able to reconstruct the input. 

The canonical Yamaguchi--Kempf protocol is therefore one concrete implementation of this abstract pattern. The purpose of the following sections is to show that the same pattern appears more generally inside QSS. Once this viewpoint is adopted, QSS architectures can be used not only to analyze encrypted cloning, but also to design new encrypted-cloning schemes.

%%%%%%%%%%%%%%%%%%%%%%%%%%%%%%
% SECTION\ 4
%

\section{Quantum secret-sharing access structures}
\label{sec:qss-access}

QSS provides the natural access-structure language for the framework developed in this paper~\cite{gottesman2000theory}. We use the isometric encoding \(V\) and the reduced channels \(\mathcal E_X\) introduced in Sec.~\ref{sec:qecl-structures}. 

\subsection{Qualified, forbidden, and intermediate sets}
In QSS language, the family of all subsets from which the input can be decoded is the \textit{access structure}. We denote it by
\begin{equation}
    \mathcal F
    =
    \left\{
        X\subseteq\Omega:
        \exists\,\mathcal D_X
        \text{ such that }
        \mathcal D_X\circ\mathcal E_X
        =
        \operatorname{id}_{\mathrm{in}}
    \right\}.
\end{equation}
The elements of \(\mathcal F\) are the \textit{qualified} sets.
This family is monotone: if \(X\in\mathcal F\) and \(X\subseteq Y\), then \(Y\in\mathcal F\). Indeed, a decoder with access to \(Y\) can ignore the shares in \(Y\setminus X\) and use the decoder for \(X\). 

A subset \(X\subseteq\Omega\) is called \emph{forbidden} if it contains no information about the secret. In channel terms, this means that there exists a fixed state \(\omega_X\), independent of the input, such that
\(\mathcal E_X(\rho)=\omega_X\)
for all \(\rho\). 

A subset that is neither qualified nor forbidden is called \emph{intermediate}. Such a subset may contain partial information about the secret, but not enough to reconstruct the full quantum state.

This gives the standard distinction between perfect and ramp QSS. In a perfect QSS scheme, every subset is either qualified or forbidden~\cite{ogawa2005quantum,zhang2015strongly,matsumoto2014coding}. There are no intermediate sets. In a ramp QSS scheme, by contrast, intermediate sets are allowed. They may leak partial information about the input while remaining unable to reconstruct it.

The quantum nature of the secret imposes an additional constraint on the access structure. Two disjoint subsets cannot both be qualified for the same unknown quantum state. Otherwise, each subset could independently reconstruct the state, producing two perfect copies and violating the no-cloning theorem. In particular, for an isometric encoding, if \(X\) is qualified then its complement \(\Omega\setminus X\) cannot also be qualified.

For \textit{pure-state encodings}, the relation between a qualified set and its complement is even stronger~\cite{cleve1999share,gottesman2000theory}. If \(X\) is \textit{qualified}, then the complementary subsystem \(\Omega\setminus X\) is \textit{forbidden}. Equivalently, recovery from \(X\) is possible exactly when the environment discarded by \(X\) is completely decoupled from the input (this is the usual erasure-correction interpretation of quantum secret sharing: reconstructing from \(X\) is the same as correcting the erasure of \(\Omega\setminus X\))~\cite{cleve1999share,knill1997theory}.

The access-structure viewpoint is therefore naturally combinatorial. The Boolean lattice \(\mathcal P(\Omega)\) organizes all possible subsets of shares. The family \(\mathcal F\) of qualified sets occupies an upper region of this lattice, while forbidden and intermediate sets occupy the non-qualified region. Perfect QSS corresponds to a sharp division between qualified and forbidden subsets. Ramp QSS allows a middle region of intermediate subsets.

This language is used in the next subsection to formulate quantum encrypted cloning as a special pattern inside \(\mathcal F\). 

%%%%%%%%%%%%%%%%%%%%%%%%%%%%%%%%
%

\subsection{QECL as a key-centered access substructure}
\label{sec:qecl-as-qss}

We can now formulate quantum encrypted cloning directly in the language of QSS access structures. Let \(\mathcal F\subseteq\mathcal P(\Omega)\) be the family of qualified sets associated with the isometric encoding introduced above. A QECL structure is not a new notion of recoverability beyond QSS. Rather, it is a special pattern inside \(\mathcal F\): a family of qualified sets sharing a common non-qualified subsystem.

More precisely, a QECL substructure consists of a subsystem \(K\subseteq\Omega\) and a family of subsystems \(S_1,\ldots,S_m\subseteq\Omega\) such that, for every \(i=1,\ldots,m\),
\begin{equation}
    K\notin\mathcal F,
    \qquad
    S_i\notin\mathcal F,
    \qquad
    K\cup S_i\in\mathcal F .
\end{equation}
The subsystem \(K\) is interpreted as the key, while the \(S_i\)'s are interpreted as encrypted clones relative to \(K\). The qualified sets \(Q_i=K\cup S_i\) are the redemption sets.

We also require \(K\) to be the common intersection of the selected redemption sets,
\begin{equation}
    K=\bigcap_{i=1}^m Q_i .
\end{equation}
This makes the structure a star-like pattern in the Boolean lattice \(\mathcal P(\Omega)\): the center \(K\) is non-qualified, each leaf \(S_i\) is non-qualified, but each union \(K\cup S_i\) is qualified.

Operationally, an encrypted clone is not a subsystem that contains a copy of the input state by itself. It is a subsystem that completes a key into a qualified set. Thus \(S_i\) is clone-like only relative to \(K\): it is non-redeeming alone, but redeeming together with the key.

A nontrivial encrypted-cloning interpretation requires at least two distinct alternatives, \(m\ge 2\). The corresponding redemption sets \(Q_i=K\cup S_i\) are different ways of recovering the same encoded input, but they are not independent copies because they all overlap on \(K\). This common overlap is what makes the structure compatible with the no-cloning theorem.

This formulation also separates non-redeemability from absence of information. The conditions \(K\notin\mathcal F\) and \(S_i\notin\mathcal F\) mean only that the key and clone subsystems are not individually qualified. They may be forbidden, as in perfect QSS, or intermediate, as in ramp QSS. In both cases the QECL interpretation remains valid as long as full recovery requires the combined system \(K\cup S_i\).

The canonical Yamaguchi--Kempf protocol fits this pattern: its signal qubits play the role of the \(S_i\)'s, while the relevant key subsystem plays the role of \(K\). More generally, any QSS access structure containing such a star-like substructure admits a QECL interpretation. This gives the extraction principle used in the next section: instead of constructing encrypted clones from scratch, one can search known QSS access structures for key-centered families of qualified sets.

%%%%%%%%%%%%%%%%%%%%%%%%%%%%%%%%%%%
% SECTION\ 5
%

\section{Exporting QSS architectures to QECL}
\label{sec:exporting-qss}

The previous section formulated quantum encrypted cloning as a key-centered substructure of a QSS access structure. This viewpoint immediately suggests a constructive principle. Instead of designing an encrypted-cloning protocol from scratch, one may start from a known QSS architecture and look for families of qualified sets with a non-qualified common intersection. Whenever such a family exists, the access structure admits an encrypted-cloning interpretation.

Let
\(\mathcal F\subseteq\mathcal P(\Omega)\)
be the family of qualified sets of a QSS scheme on the set of shares \(\Omega\). The general task is to identify subsystems
\(K,S_1,\ldots,S_m\subseteq\Omega\)
such that
\begin{equation}
K\notin\mathcal F,
    \qquad
    S_i\notin\mathcal F,
    \qquad
    K\cup S_i\in\mathcal F
\end{equation}
for every \(i=1,\ldots,m\). The sets
\(Q_i=K\cup S_i\)
are then the redemption sets. The common subsystem \(K\) is interpreted as the key, while the \(S_i\)'s are interpreted as encrypted clones relative to \(K\).

This section develops this extraction principle for several classes of QSS architectures. The examples are not meant to exhaust the possible realizations of each class, nor to reproduce the most general QSS construction of that type. Rather, they show how the corresponding access-structure pattern can be realized by simple isometries and then reinterpreted as an encrypted-cloning architecture.

\subsection{General extraction rule}
\label{subsec:general-extraction-rule}

Given an access structure \(\mathcal F\), one searches the Boolean lattice \(\mathcal P(\Omega)\) for a star-like pattern centered on a non-qualified subsystem \(K\). The leaves of the star are non-qualified subsystems \(S_i\), while the unions \(K\cup S_i\) are qualified.

\newpage
Equivalently, 
\begin{itemize}
\item start from a family of qualified sets
\(Q_1,\ldots,Q_m\in\mathcal F\)
\item define
\(K=\bigcap_{i=1}^m Q_i.\)
\item if \(K\notin\mathcal F\), then \(K\) is a candidate key. 
\item in this case set
\(S_i=Q_i\setminus K\)
and obtain
\(Q_i=K\cup S_i.\)
\item
If each \(S_i\) is also non-qualified, i.e., \(\forall i: S_i\notin\mathcal F\), then the family defines a QECL substructure.
\end{itemize}

We will illustrate this procedure with a few examples. A particularly simple class of examples will be provided by coherent syndrome encodings (see Section \ref{subsubsec:duplicated-syndrome-prototype}): these will make the key--clone structure explicit while using only elementary controlled-unitary operations.
Then, 
we are going to look at threshold schemes, ramp schemes, hierarchical schemes and compartmented schemes.

Importantly, those QSS categories should not be understood as mutually exclusive.
They emphasize different aspects of a QSS architecture: 
a threshold scheme is characterized by the cardinality of its qualified sets; a ramp scheme by the presence of intermediate, partially informative non-qualified sets; a hierarchical scheme by the existence of distinguished classes of shares; and a compartmented scheme by constraints on the types of shares that must be combined. 
The same encoding may therefore instantiate more than one of these viewpoints. The purpose of this section is not to classify schemes into disjoint families, but to show how different QSS design principles can be reinterpreted as QECL architectures.

%%%%%%%%%%%%%%
\subsubsection{Prototype: coherent duplicated-syndrome encoding}
\label{subsubsec:duplicated-syndrome-prototype}

Consider a key qubit \(A\) and two signal qubits \(S_1,S_2\). 
The input is an arbitrary qubit state
\(
    \ket{\psi}_A
    =
    \alpha\ket{0}_A+\beta\ket{1}_A .
\)
We append to \(A\) the Bell state
\begin{equation}
\ket{\Phi^+}_{S_1S_2}
    =
    \frac{1}{\sqrt{2}}
    \left(
        \ket{00}_{S_1S_2}
        +
        \ket{11}_{S_1S_2}
    \right).
\end{equation}
A minimal coherent encoding is obtained as follows. We apply a controlled-\(Z\) gate from any of the signal qubits (they are all equal). Choosing \(S_1\), the encoding isometry is $V=\operatorname{CZ}_{S_1\to A}\otimes \mathbb{I}_{S_2}$, with
\begin{equation}
\operatorname{CZ}_{S_1\to A}
    =
    \ket{0}\!\bra{0}_{S_1}\otimes I_A
    +
    \ket{1}\!\bra{1}_{S_1}\otimes Z_A,
\end{equation}
which yields
\begin{equation}
V\ket{\psi}
    =
    \frac{1}{\sqrt{2}}
    \left(
        \ket{\psi}_A\ket{00}_{S_1S_2}
        +
        Z\ket{\psi}_A\ket{11}_{S_1S_2}
    \right).
\end{equation}
The encoding may be viewed as a GHZ-controlled Pauli encryption. The signal register stores, coherently and \textit{redundantly}, a binary decryption syndrome. If the syndrome is \(0\), the key qubit \(A\) carries \(\ket{\psi}\). If the syndrome is \(1\), the key qubit carries \(Z\ket{\psi}\). 
Each individual signal qubit \(S_i\) contains enough syndrome information to decide which correction must be applied to \(A\). For instance, if the chosen redemption set is \(AS_2\), the decoder simply applies a second controlled-\(Z\),
\(D_{AS_2} = \operatorname{CZ}_{S_2\to A}.\)
Indeed,
\begin{equation}
\operatorname{CZ}_{S_2\to A}
    V\ket{\psi}
    =
    \ket{\psi}_A
    \otimes
    \ket{\Phi^+}_{S_1S_2}.
\end{equation}
Thus \(AS_2\) is qualified: access to \(S_2\) is enough to refocus the encoded state back onto the key qubit \(A\).
Thus
\(
    AS_i
\)
is qualified for every \(i\).

On the other hand, \(S_i\) alone is not qualified. Its reduced state is
\(
    \rho_{S_i}
    =
    {I}\big/{2},
\)
independent of \(\ket{\psi}\). 

The key qubit \(A\) alone is also not qualified. Its reduced state is
\begin{equation}
\rho_A
    =
    \frac{1}{2}
    \left(
        \ket{\psi}\!\bra{\psi}
        +
        Z\ket{\psi}\!\bra{\psi}Z
    \right)=
    \frac{1}{2}
    \left(
        I+r_zZ
    \right),
\end{equation}
where $r_z$ is the $z$-component of the Bloch-vector 
\(
    \ket{\psi}\!\bra{\psi}
    =
    \frac{1}{2}
    \left(
        I+r_xX+r_yY+r_zZ
    \right)
\).
Thus \(A\) is intermediate. 

It is easy to check that the set $S_1S_2$ is intermediate: only information from the Bloch channel $z$ is available to the pair
\begin{equation}
\rho_{S_1 S_2}=\frac{1}{2}\left[|00\rangle\langle 00|+|11\rangle\langle 11|+r_z(|00\rangle\langle 11|+|11\rangle\langle 00|)\right] = \frac{1}{2}\left(I+r_z X\right)
\end{equation}
Therefore the access pattern of this prototype protocol is
\begin{equation}
A\notin\mathcal F,
    \qquad
    S_1\notin\mathcal F,
    \qquad 
    S_2\notin\mathcal F,
    \qquad
    S_1S_2\notin\mathcal F,
    \qquad
    AS_1\in\mathcal F
        \qquad
        AS_2\in\mathcal F
\end{equation}
In the QECL interpretation, \(A\) is the key and the signal qubits \(S_i\) are encrypted clones. Each \(S_i\) completes the key into a qualified redemption set, while neither the key nor any signal qubit is qualified on its own.

The \(m\)-signal version is obtained by replacing the Bell state by the GHZ state
\begin{equation}
\ket{\mathrm{GHZ}_m}_{S_1\cdots S_m}
    =
    \frac{1}{\sqrt{2}}
    \left(
        \ket{0}^{\otimes m}
        +
        \ket{1}^{\otimes m}
    \right)
\end{equation}
and applying \(\operatorname{CZ}_{S_i\to A}\) from any one of the signal qubits, since all \(S_i\)'s are perfectly correlated on the GHZ support.
In this case one can verify that the whole signal register
\(S_1\cdots S_m\)
is not qualified. Tracing out \(A\), we obtain
\begin{equation}
\rho_{S_1\cdots S_m}
    =
    \frac{1}{2}
    \left[
        \ket{0}^{\otimes m}\!\bra{0}^{\otimes m}
        +
        \ket{1}^{\otimes m}\!\bra{1}^{\otimes m}
        +
        r_z
        \left(
            \ket{0}^{\otimes m}\!\bra{1}^{\otimes m}
            +
            \ket{1}^{\otimes m}\!\bra{0}^{\otimes m}
        \right)
    \right],
\end{equation}
where \(r_z\) is the \(z\)-component of the input Bloch vector. Therefore the entire signal register depends only on \(r_z\), and is independent of the transverse components \(r_x\) and \(r_y\). Since different input states with the same \(r_z\) but different \(r_x,r_y\) induce the same reduced state on \(S_1\cdots S_m\), no decoding channel acting only on \(S_1\cdots S_m\) can reconstruct an arbitrary input qubit. Hence
\(S_1\cdots S_m\notin\mathcal F .\)

\subsection{QECL\ schemes from threshold architectures}
\label{subsec:threshold-qecl}

Threshold QSS schemes provide the simplest access-structure source of QECL patterns. In a \((t,n)\)-threshold scheme, every subset of at least \(t\) shares is qualified, while subsets below the threshold are not qualified. Thus the family of qualified sets is determined only by cardinality:
\begin{equation}
X\in\mathcal F
    \quad\Longleftrightarrow\quad
    |X|\ge t .
\end{equation}
This immediately produces a key-centered QECL substructure. Choose a subsystem \(K\subseteq\Omega\) of size
\(|K|=t-1 .\)
Then \(K\notin\mathcal F\). However, for every share \(s\in\Omega\setminus K\),
\(K\cup\{s\}\in\mathcal F ,\)
because
\(|K\cup\{s\}|=t .\)
Thus every share outside \(K\) completes the key into a qualified set. In the QECL interpretation, \(K\) is the key and the single-share systems
\(S_s=\{s\}\)
are encrypted clones relative to \(K\). The redemption sets are
\(Q_s=K\cup\{s\}.\)

This shows that threshold access structures naturally generate symmetric QECL patterns. The key has one share fewer than the reconstruction threshold, and every additional share can be used as an alternative encrypted clone. The resulting structure is symmetric because all shares outside \(K\) play the same access-structural role.

\subsubsection{Example: a \((2,m+1)\)-threshold pattern}
\label{subsubsec:threshold-example}

In the secret sharing $(k,m+1)$-threshold patterns there are $m+1$ shares and $k$ are needed for the reconstruction.
The prototype of Sec.~\ref{subsubsec:duplicated-syndrome-prototype} realizes a threshold-like QECL pattern with $k=2$ (although not a full perfect threshold QSS scheme.). Let
\(\Omega=\{A,S_1,\ldots,S_m\},\)
and consider the coherent encoding
\begin{equation}
V\ket{\psi}
    =
    \frac{1}{\sqrt{2}}
    \left(
        \ket{\psi}_A\ket{0}^{\otimes m}_{S_1\cdots S_m}
        +
        Z\ket{\psi}_A\ket{1}^{\otimes m}_{S_1\cdots S_m}
    \right).
\end{equation}
For every \(i\), the reduced state on \(AS_i\) is
\begin{equation}
\rho_{AS_i}
    =
    \frac{1}{2}
    \ket{\psi}\!\bra{\psi}_A
    \otimes
    \ket{0}\!\bra{0}_{S_i}
    +
    \frac{1}{2}
    Z\ket{\psi}\!\bra{\psi}Z_A
    \otimes
    \ket{1}\!\bra{1}_{S_i}.
\end{equation}
Hence \(AS_i\) is qualified: applying \(\operatorname{CZ}\) to \(A\) conditioned on \(S_i=1\) refocuses the state onto \(A\).

In threshold language, the minimal qualified sets of the QECL substructure are
\(AS_1,\ AS_2,\ \ldots,\ AS_m .\)
Thus the key is
\(
    K=A,
\)
and the encrypted clones are the individual signal qubits
\(
    S_1,\ldots,S_m.
\)
Each \(S_i\) completes \(A\) into a qualified set:
\(
    A\cup S_i\in\mathcal F.
\)
Neither \(A\) nor any \(S_i\) is qualified by itself:
\(
    A\notin\mathcal F,
    \;
    S_i\notin\mathcal F
\).

This example realizes the access pattern of a \((2,m+1)\)-threshold structure restricted to the star centered on \(A\) (it should not be confused with a full perfect \((2,m+1)\)-threshold QSS scheme, since, as we have seen, other subsets may be intermediate rather than forbidden).

The point of the example is that the threshold mechanism already contains the QECL idea: a subsystem just below the reconstruction threshold becomes a key, and any one additional share becomes an encrypted clone relative to that key.

%%%%%%%%%%%%%%%%%%%%%%%%%%%%%%%%

\subsection{QECL\ schemes from ramp architectures}
\label{subsec:ramp-qecl}

Ramp QSS schemes provide a natural source of leaky encrypted-cloning structures~\cite{ogawa2005quantum}. In a perfect QSS scheme, every non-qualified subset is forbidden: it contains no information about the secret. In a ramp QSS scheme, by contrast, there may be intermediate subsets. These subsets are not qualified, hence they cannot reconstruct the secret, but they may nevertheless contain partial information about it.

This distinction is essential for QECL. The encrypted-cloning condition requires that a signal subsystem be non-qualified by itself and become qualified when combined with a key. 
Thus ramp QSS naturally supports QECL architectures in which some non-redeeming subsystems are leaky but still unable to reconstruct the secret.
The prototype protocol of the previous subsection already provides an example of this kind of QSS/QECL pattern. Another one follows.

This distinction is not merely formal. 
The canonical Yamaguchi--Kempf protocol itself can be read in this way whenever non-redeeming subsystems retain partial information about the input. The Cear\u{a} protocol~\cite{ceara2026cloningencryptedquantumstates} extending the original Yamaguchi and Kempf protocol to $d>2$ dimensions, has been shown to display leakage patterns~\cite{bai2026classification}, thus it also can be considered ramp.

In such cases, the protocol still realizes encrypted cloning, because suitable key--signal pairs can redeem the state, but it should not be identified with a perfect QSS scheme. 
Rather, its access-structure interpretation is ramp-like: some non-qualified subsystems are intermediate instead of forbidden. 
This illustrates why the ramp viewpoint is essential for QECL. 
It separates the redemption requirement from the stronger no-leakage requirement imposed by perfect secret sharing.

\subsubsection{Example: the \([[4,2,2]]\) ramp code}
\label{subsubsec:ramp-422-example}

A standard minimal example is provided by the \([[4,2,2]]\) stabilizer code~\cite{choi2013entanglement,matsumoto2017unitary,matsumoto2020classical}.
The \([[n,k,d]]\) codes encode $k$ logical qubits into $n$ physical qubits and can correct up to $
\lfloor(d-1) / 2\rfloor
$ arbitrary errors and up to $d-1$ known erasures.  
The following code encodes $k=2$ logical qubits into $n=4$ physical qubits and can correct up to $d-1=1$ erasure of physical qubit. 
In the secret-sharing interpretation, the $4$ physical qubits are the shares, and the $2$ logical qubits form the quantum secret.

Let the four physical shares be
\(\Omega=\{1,2,3,4\}.\)
The stabilizer is generated by
\(XXXX, \qquad ZZZZ.\)
A convenient logical basis is
\begin{equation}
\begin{aligned}
    \ket{00_L}
    &=
    \frac{1}{\sqrt{2}}
    \left(
        \ket{0000}+\ket{1111}
    \right),\\
    \ket{01_L}
    &=
    \frac{1}{\sqrt{2}}
    \left(
        \ket{0011}+\ket{1100}
    \right),\\
    \ket{10_L}
    &=
    \frac{1}{\sqrt{2}}
    \left(
        \ket{0101}+\ket{1010}
    \right),\\
    \ket{11_L}
    &=
    \frac{1}{\sqrt{2}}
    \left(
        \ket{0110}+\ket{1001}
    \right).
\end{aligned}
\end{equation}
Thus an arbitrary two-qubit secret
\begin{equation}
\ket{\psi}
    =
    \alpha\ket{00}
    +
    \beta\ket{01}
    +
    \gamma\ket{10}
    +
    \delta\ket{11}
\end{equation}
is encoded as
\begin{equation}
\begin{aligned}
    V\ket{\psi}
    =
    \frac{1}{\sqrt{2}}
    \big(
    &\alpha(\ket{0000}+\ket{1111})
    +
    \beta(\ket{0011}+\ket{1100})\\
    &+
    \gamma(\ket{0101}+\ket{1010})
    +
    \delta(\ket{0110}+\ket{1001})
    \big).
\end{aligned}
\end{equation}
The access structure is ramp-like~\cite{matsumoto2017unitary,matsumoto2020classical}. Since the code corrects the erasure of any one physical qubit, every set of three shares is qualified:
\begin{equation}
\{1,2,3\},
    \quad
    \{1,2,4\},
    \quad
    \{1,3,4\},
    \quad
    \{2,3,4\}
    \in\mathcal F .
\end{equation}
Equivalently, this can be seen from the complementary single-share reductions. For every physical qubit \(i\), the reduced state is independent of the logical input: one can show that the state is maximally mixed
\begin{equation}
\rho_i
    =
    \operatorname{Tr}_{\Omega\setminus\{i\}}
    \left(
        V\rho_L V^\dagger
    \right)
    =
    \frac{I_i}{2},
    \qquad
    i=1,2,3,4.
\end{equation}
Thus every single share is forbidden. Since the encoding is pure, the complement of a forbidden subsystem is qualified. Therefore every triple
\(\Omega\setminus\{i\}\)
is qualified.

On the other hand, pairs of shares are intermediate. They are not qualified, but they are not forbidden either. To see this explicitly, consider the reduced state on shares \(1,2\). Write the two-qubit logical secret as
\begin{equation}
\rho_L
    =
    \frac{1}{4}
    \sum_{\mu,\nu\in\{I,X,Y,Z\}}
    r_{\mu\nu}\,
    \sigma_\mu\otimes\sigma_\nu,
\end{equation}
with
\begin{equation}
r_{\mu\nu}
    =
    \operatorname{Tr}
    \left[
        \rho_L
        \left(
            \sigma_\mu\otimes\sigma_\nu
        \right)
    \right].
\end{equation}
For the pair \(\{1,2\}\), one obtains
\begin{equation}
\rho_{12}
    =
    \frac{1}{4}
    \left(
        II
        +
        r_{ZI}\,ZZ
        +
        r_{IX}\,XX
        -
        r_{ZX}\,YY
    \right).
\end{equation}
Thus \(\{1,2\}\) is not forbidden, since its reduced state depends on the logical input through the components
\(r_{ZI}, \qquad r_{IX}, \qquad r_{ZX}.\)
However, it is not qualified either: a general two-qubit secret has fifteen independent Bloch components, while \(\rho_{12}\) contains only three of them. Hence \(\{1,2\}\) is an intermediate set. By symmetry, the same conclusion holds for every pair of physical shares.

By contrast, every triple of physical shares is qualified, because its complement is maximally mixed: by construction a triple contains a complete representation of the encoded logical algebra and therefore suffices to recover all fifteen components of the secret.

We may now extract a QECL substructure. Choose, for instance,
\(K=\{1,2\}.\)
This pair is not qualified:
\(\{1,2\}\notin\mathcal F .\)
Now take the two candidate encrypted clones
\(S_3=\{3\}, \qquad S_4=\{4\}.\)
Individually, they are forbidden and hence not qualified:
\begin{equation}
\rho_3=\rho_4=\frac{I}{2},
    \qquad
    \{3\},\{4\}\notin\mathcal F .
\end{equation}
However,
\(K\cup S_3=\{1,2,3\}\in\mathcal F,\)
and
\(K\cup S_4=\{1,2,4\}\in\mathcal F.\)
Thus the pair \(K=\{1,2\}\) acts as a key subsystem, while the single shares \(3\) and \(4\) act as encrypted clones relative to that key. Either \(K\cup S_3\) or \(K\cup S_4\) can reconstruct the two-qubit secret.

The important point is that the key \(K=\{1,2\}\) is intermediate rather than forbidden. It contains partial information about the logical secret, as shown by \(\rho_{12}\), but it cannot reconstruct the full two-qubit state. This is precisely the ramp feature. In the QECL interpretation, such leakage is allowed: the key and the encrypted clones need only be non-qualified individually, while their unions must be qualified.

%%%

This example shows how a genuine ramp QSS code can be reinterpreted as a QECL architecture. The resulting encrypted clones are not produced by the duplicated-syndrome prototype. Rather, they arise from the erasure-correction structure of a stabilizer ramp code. The access-structure pattern is
\begin{equation}
\{1,2\}\notin\mathcal F,
    \qquad
    \{3\},\{4\}\notin\mathcal F,
    \qquad
    \{1,2,3\},\{1,2,4\}\in\mathcal F.
\end{equation}
Thus the code contains a key-centered family of qualified sets and therefore supports a QECL interpretation.

More generally, any ramp QSS scheme in which intermediate sets can be completed into qualified sets by adding one of several alternative shares gives rise to the same kind of QECL structure. The ramp feature changes the security interpretation of the non-qualified sets, but not the redemption logic: the encrypted clones are precisely those subsystems that complete a non-qualified key into a qualified set.

The \([[4,2,2]]\) example also shows why stabilizer constructions are useful for QECL. 
They do not define a separate logical type of access structure; rather, they provide an implementable algebraic realization of threshold, ramp, or other access patterns.

\paragraph{Stabilizer and graph-state realizations.}
Notice that stabilizer based (and graph-state base) QSS schemes should be viewed as an implementation layer for the access-structure patterns discussed above. 
They may realize threshold, ramp, hierarchical, or more general access structures, while providing explicit encodings, logical operators, and decoding circuits. 
For QECL, this is particularly useful because the extraction rule can be applied directly to the qualified sets of the stabilizer code: once overlapping qualified supports with common non-qualified intersection are identified, the corresponding key--clone interpretation follows.

%%%%%%%%%%%%%%%%%%%%%%%%%%%%
%

\subsection{QECL\ schemes from hierarchical architectures}
\label{subsec:hierarchical-qecl}

Hierarchical QSS schemes distinguish different classes of participants~\cite{tassa2007hierarchical,lai2022dynamic}. Some shares have a privileged role in the access structure: they can reconstruct the secret with fewer additional shares than lower-level participants. This is particularly natural from the QECL viewpoint, because encrypted cloning already distinguishes between a key subsystem and subsystems that complete the key into a qualified set.

A typical hierarchical pattern has one high-level participant \(H\) and several low-level participants
\(L_1,\ldots,L_m .\)
The high-level share is not qualified by itself, and no low-level share is qualified by itself, but each pair
\(H L_i\)
is qualified. At the same time, pairs of low-level shares such as
\(L_i L_j\)
may remain non-qualified. Thus the access structure is not symmetric: the role of \(H\) is structurally different from the role of the \(L_i\)'s.

In QECL language, this gives a privileged-key architecture. The subsystem \(H\) is interpreted as the key, and the lower-level shares \(L_i\) are encrypted clones relative to \(H\):
\begin{equation}
H\notin\mathcal F,
    \qquad
    L_i\notin\mathcal F,
    \qquad
    H\cup L_i\in\mathcal F .
\end{equation}
The hierarchy is reflected in the fact that the \(L_i\)'s can complete \(H\), whereas they cannot generally complete each other.

The hierarchical viewpoint is already visible in the simple constructions discussed above. In the duplicated-syndrome prototype of Section \ref{subsubsec:duplicated-syndrome-prototype} one may take 
\(H=A, \; L_i=S_i .\)
The access pattern is hierarchical because the roles of \(A\) and the \(S_i\)'s are not interchangeable: the \(S_i\)'s are alternative completions of the key, whereas \(A\) is the common subsystem shared by all redemption sets.

The same interpretation applies to the canonical Yamaguchi--Kempf protocol with
\(H=K, \), where we denote by
\(K\)
the key block required for decryption, and $L_i=S_i$: the redemption sets have the form
\(K S_i .\)
In this sense, the canonical QECL protocol already exhibits a hierarchical organization. Being an extension of the original Yamaguchi and Kempf protocol to $d>2$ dimensions, also the Cear\u{a} protocol~\cite{ceara2026cloningencryptedquantumstates} can be considered hierarchical.

The next example shows that this hierarchical interpretation is not restricted to the duplicated-syndrome or Yamaguchi--Kempf constructions.

\subsubsection{Example: asymmetric qubit syndrome splitting}
\label{subsubsec:hierarchical-qubit-example}

A simple qubit realization of a hierarchical QECL pattern can be obtained by distributing the decryption syndrome unevenly among lower-level subsystems. Let \(H\) be the high-level subsystem, interpreted as the key. We introduce a strong lower-level subsystem
\[
    S=S_aS_b
\]
made of two qubits, and two additional lower-level qubits \(W\) and \(M\). We use two binary syndrome variables
\(
    a,b\in\{0,1\}
\)
and define the Pauli encryption
\[
    U_{ab}=Z^aX^b .
\]
The encoding isometry is
\begin{equation}
    V_{\mathrm{hier}}\ket{\psi}
    =
    \frac{1}{2}
    \sum_{a,b\in\{0,1\}}
    U_{ab}\ket{\psi}_H
    \otimes
    \ket{ab}_{S_a S_b}
    \otimes
    \ket{a}_{W}
    \otimes
    \ket{b}_{M}.
\end{equation}
The subsystem \(S=S_aS_b\) contains the full syndrome \((a,b)\). The weak subsystem \(W\), by contrast, contains only the first syndrome component \(a\), while \(M\) contains the second component \(b\).

Thus, the set
\[
    HS
    =
    H S_aS_b
\]
is qualified. Indeed, access to \(S_aS_b\) determines both syndrome bits and hence the correction \(U_{ab}^{\dagger}\). The corresponding decoder is the controlled unitary
\[
    D_{HS}
    =
    \sum_{a,b\in\{0,1\}}
    \ket{a}\!\bra{a}_{S_a}
    \otimes
    \ket{b}\!\bra{b}_{S_b}
    \otimes
    X_H^b Z_H^a .
\]
Applying \(D_{HS}\) to the reduced state on \(HS\) recovers the input state on \(H\).

The weak subsystem \(W\) is not sufficient by itself to complete the key, since \(HW\) lacks the second syndrome bit \(b\). However, \(W\) can be completed by the auxiliary subsystem \(M\). The set
\[
    HWM
\]
is qualified, because \(W\) and \(M\) together determine the full syndrome \((a,b)\). The corresponding decoder is
\[
    D_{HWM}
    =
    \sum_{a,b\in\{0,1\}}
    \ket{a}\!\bra{a}_{W}
    \otimes
    \ket{b}\!\bra{b}_{M}
    \otimes
    X_H^b Z_H^a .
\]
Thus \(D_{HWM}\) recovers the input by applying \(U_{ab}^{\dagger}\) to \(H\), conditioned on the values stored in \(W\) and \(M\). The minimal redemption sets displayed by this construction are therefore
\[
    Q_1=HS,
    \qquad
    Q_2=HWM .
\]
For the purpose of QECL reformulation, their common intersection is
\[
    K=Q_1\cap Q_2=H .
\]
i.e., the key is \(K=H\), while the corresponding encrypted clones are
\[
    S_1=S=S_aS_b,
    \qquad
    S_2=WM .
\]
The first is concentrated in the strong lower-level subsystem \(S\), while the second is composite and contains the weak clone component \(W\) together with the auxiliary component \(M\).
In this way, the same key \(H\) admits inequivalent redemption modes $HS$ and $HWM$, with different lower-level resources.

The non-qualified sets reflect the same hierarchy. The key alone is not qualified:
\(
    H\notin\mathcal F 
\);
the strong subsystem alone is not qualified:
\(
    S\notin\mathcal F 
\);
the weak component is also not qualified:
\(
    W\notin\mathcal F
\);
moreover,
\(
    HW\notin\mathcal F
\), and
\(
    HM\notin\mathcal F
\),
because each of these sets lacks one syndrome component. By contrast,
\(
    HS\in\mathcal F,
    \qquad
    HWM\in\mathcal F
\).

%%%%%%%%%%%%%%%%%%%%%%%%

\subsection{QECL\ schemes from compartmented architectures}
\label{subsec:compartmented-qecl}

Compartmented QSS schemes divide the set of shares into distinct classes or compartments \cite{brickell1989some,chen2024ideal}. A subset is qualified only if it contains appropriate contributions from the required compartments. This produces access structures that are neither purely threshold nor necessarily hierarchical: the relevant feature is not only the number of shares, but also their type.

This viewpoint is useful for QECL because it shows that encrypted clones need not be individual shares. In a compartmented architecture, the subsystem that completes a key may itself be composite. For example, a key \(K\) may become qualified only when combined with one share from compartment \(X\) and one share from compartment \(Y\). In that case, the encrypted clones are not \(X_i\) or \(Y_j\) separately, but the composite systems
\(S_{ij}=X_iY_j .\)
The corresponding redemption sets have the form
\(Q_{ij}=K\cup X_i\cup Y_j.\)

Thus compartmented QSS naturally leads to encrypted clones with internal structure. This is useful when different physical or organizational resources are required for redemption. One may think of the key as a central subsystem, while each encrypted clone is assembled from one component of each required compartment.

A compartmented pattern is also visible inside the canonical Yamaguchi--Kempf protocol, although it does not exhaust its full access structure. Besides the larger key--signal redemption sets discussed above; as shown in \cite{gianini2026full}  one may consider the smaller qualified subsets consisting of the input/output register \(A\), one signal qubit \(S_i\), and its corresponding key qubit \(K_i\). These sets have the form
\(A S_i K_i .\)
They are compartmented in the sense that reconstruction requires one subsystem from each of three roles: the register \(A\), a signal subsystem, and a key subsystem. Relative to the schematic form
\(Q_{ij}=K\cup X_i\cup Y_j ,\)
the Yamaguchi--Kempf pattern corresponds to the constrained case in which the signal and key labels must match:
\(Q_i = A\cup S_i\cup K_i .\)
Thus only the diagonal combinations \(i=j\) are singled out by this particular compartmented substructure. Other authorized sets may exist in the full Yamaguchi--Kempf access structure, so this should be understood as a compartmented substructure rather than as a complete characterization of the protocol.

\subsubsection{Example: split-syndrome compartmented encoding}
\label{subsubsec:compartmented-example}

Consider a key qubit \(A\) and two compartments
\(X=\{X_1,X_2\}, \qquad Y=\{Y_1,Y_2\}.\)
We construct an encoding in which the correction syndrome has two binary components
\((a,b)\in\{0,1\}^2.\)
The \(X\)-compartment redundantly stores the first component \(a\), while the \(Y\)-compartment redundantly stores the second component \(b\). Let
\(U_{ab}=Z^a X^b,\)
and define the isometry
\begin{equation}
V_{\mathrm{comp}}\ket{\psi}
    =
    \frac{1}{2}
    \sum_{a,b\in\{0,1\}}
    U_{ab}\ket{\psi}_A
    \otimes
        \ket{aa}_{X_1 X_2}
    \otimes
        \ket{bb}_{Y_1 Y_2}.
\end{equation}
The four syndrome branches are mutually orthogonal, and the corresponding corrections \(U_{ab}\) are unitary. Hence \(V_{\mathrm{comp}}\) is an isometric encoding.

The access structure is compartmented. A subsystem of the form
\(A X_iY_j\)
is qualified. Indeed, \(X_i\) reveals the value of \(a\), while \(Y_j\) reveals the value of \(b\). Together with the key qubit \(A\), they determine the correction
\(U_{ab}^{\dagger}\)
needed to refocus the state onto \(A\). Thus
\(A X_iY_j\in\mathcal F\)
for all
\(i,j\in\{1,2\}.\)

By contrast, \(AX_i\) alone is not qualified: it identifies \(a\) but not \(b\). Similarly, \(AY_j\) alone is not qualified: it identifies \(b\) but not \(a\). Therefore a full clone relative to \(A\) must contain one component from each compartment. The encrypted clones are the composite systems
\(S_{ij}=X_iY_j,\)
and the redemption sets are
\(Q_{ij}=A X_iY_j.\)
The QECL pattern is
\begin{equation}
A\notin\mathcal F,
    \qquad
    X_iY_j\notin\mathcal F,
    \qquad
    A X_iY_j\in\mathcal F.
\end{equation}
This example differs from the duplicated-syndrome prototype in an important way. In the prototype, each individual signal qubit carries the entire decryption syndrome. Here the syndrome is split across compartments. No single \(X_i\) or \(Y_j\) is an encrypted clone by itself. Rather, an encrypted clone is a composite subsystem \(X_iY_j\), assembled from the required compartments.

The reduced state of an individual compartment share is independent of the input:
\(\rho_{X_i} = \rho_{Y_j} = \frac{I}{2}.\)
However, the key \(A\) alone is maximally mixed:
\begin{equation}
\rho_A
    =
    \frac{1}{4}
    \sum_{a,b\in\{0,1\}}
    U_{ab}\ket{\psi}\!\bra{\psi}U_{ab}^{\dagger}
    =
    \frac{I}{2}.
\end{equation}
Thus the key is forbidden in this particular example. The composite clone \(X_iY_j\) is also non-qualified by itself: it contains the classical syndrome \((a,b)\), but no quantum system carrying the encrypted state. Only when \(X_iY_j\) is combined with \(A\) can the syndrome be used to undo the Pauli encryption.

The compartmented interpretation is therefore clear:
\begin{equation}
K=A,
    \qquad
    S_{ij}=X_iY_j,
    \qquad
    Q_{ij}=A X_iY_j.
\end{equation}
There are four alternative redemption sets,
\begin{equation}
AX_1Y_1,\quad AX_1Y_2,\quad AX_2Y_1,\quad AX_2Y_2,
\end{equation}
all sharing the same key \(A\). The different encrypted clones are composite and internally structured, reflecting the compartmented nature of the underlying access structure.

More generally, one may split the decryption syndrome into several components and distribute each component redundantly within a different compartment. A valid encrypted clone is then any subsystem containing enough components to reconstruct the full syndrome. This gives a systematic way to build QECL schemes in which the clones are not individual shares, but structured assemblies of shares from different classes.

\begin{table}[t]
\centering
\caption{Summary of the QSS-to-QECL examples discussed in this work.}
\label{tab:qss-qecl-summary}
\renewcommand{\arraystretch}{1.25}
\begin{tabularx}{\linewidth}{L{0.19\linewidth} Y Y Y}
\hline
\textbf{Architecture} & 
\textbf{Key \(K\)} & 
\textbf{Encrypted clones \(S_i\)} & 
\textbf{Redemption sets / interpretation} \\
\hline

Duplicated-syndrome construction &
Common syndrome subsystem \(A\) &
Signal subsystems \(S_i\) &
Each pair \(A\cup S_i\) is qualified and acts as a valid redemption set. \\

Threshold-like construction &
A fixed set of \(t-1\) shares &
Individual additional shares \(s_i\) &
Each \(K\cup\{s_i\}\) reaches threshold size \(t\), yielding a threshold-like QECL pattern. \\

Ramp-code construction &
A non-qualified intermediate subsystem, such as \(\{1,2\}\) in the \([[4,2,2]]\) example &
Additional shares, such as \(\{3\}\) and \(\{4\}\) &
The sets \(\{1,2,3\}\) and \(\{1,2,4\}\) are qualified, while leakage below full qualification is allowed. \\

Hierarchical construction &
High-level share or high-level block \(H\) &
Lower-level subsystems, such as \(S\) or \(W\cup M\) &
Authorized combinations such as \(H\cup S\) or \(H\cup W\cup M\) instantiate QECL redemption sets. \\

Compartmented construction &
A distinguished compartment share \(A\) &
Cross-compartment subsystems \(X_i\cup Y_j\) &
Sets of the form \(A\cup X_i\cup Y_j\) realize compartment-constrained redemption. \\

\hline
\end{tabularx}
\end{table}

%%%%%%%%%%%%%%%%%%%%%%%% 
% SEC 6

\section{Discussion and outlook}
\label{sec:discussion}

The main point of this work is that quantum encrypted cloning does not require a new notion of recoverability beyond quantum secret sharing. Its distinctive content lies instead in an operational reinterpretation of selected access structures. Whenever a family of qualified sets has a non-qualified common intersection, that intersection can be read as a key, and the non-common parts can be read as encrypted clones relative to that key. In this sense, QECL is not external to QSS: it is a key-centered way of reading overlapping qualified sets.

This viewpoint also clarifies the role of leakage. Perfect QSS gives rise to a strong form of encrypted cloning, in which non-qualified subsystems are forbidden. However, encrypted cloning as such only requires non-redeemability of the key and clone subsystems taken separately, together with redeemability of their union. Ramp QSS is therefore a natural setting for QECL, because intermediate subsystems may carry partial information without being able to reconstruct the secret. Leakage may rule out a perfect-secret-sharing interpretation, but it does not by itself invalidate encrypted cloning.

The examples discussed above illustrate that this is not only a formal observation. Threshold-like architectures give a direct way to obtain symmetric key--clone redemption sets. Ramp codes, such as the \([[4,2,2]]\) example, show how leaky non-qualified keys can still participate in valid QECL structures. Hierarchical architectures emphasize the privileged role of a key subsystem, while compartmented architectures show that encrypted clones need not be individual shares: they may be composite systems assembled from different classes of shares. These examples suggest that QSS provides a library of architectural motifs for encrypted quantum redundancy.

The same picture can be translated into the language of quantum error correction~\cite{knill1997theory,nielsen2010quantum}. For an isometric encoding, a qualified set is a region from which the input can be recovered, equivalently a region whose complement is a correctable erasure. Thus QECL structures may also be seen as families of overlapping erasure-recovery regions with a common subsystem. This error-correction formulation is useful, especially when working with stabilizer or graph-state realizations, because qualified sets can often be characterized through logical-operator support or erasure-correction conditions. Nevertheless, the access-structure language remains the most direct one for the present purpose.

Recent work \cite{lim2026encrypted} has related canonical QECL protocols to Absolute Maximal Entanglement type structures and mapped specific instances of the protocol to threshold QSS.\footnote{Specifically, they show that $d$-dimensional Bell state under partial encryption, where only a single register is encrypted, there exist an encrypted cloning protocol with $n=2$ signal-noise qudit pairs (up to a local unitary) that exhibits a one-to-one correspondence to a pure state $((3,5))$ threshold QSS scheme with secret and share dimensions $d$, where $d \geq 2$.}

The perspective developed in the present work is complementary. Rather than starting from a given encrypted-cloning protocol and mapping it into QSS, we start from QSS access structures and ask when they generate a QECL interpretation. This reversal is important because it separates the redemption pattern from the stronger requirements of perfect threshold secret sharing. In particular, when non-redeeming subsystems retain partial information, the appropriate interpretation is ramp-like rather than perfect-threshold.

What was less explicit was the operational reading of these ingredients as delayed and alternative redemption opportunities. A subsystem that completes a key into a qualified set was usually viewed simply as part of an authorized set or a recovery region. The QECL viewpoint singles out the same structure as an encrypted clone relative to a key.

Several directions remain open. One is the classification problem: given a QSS access structure, characterize all key-centered QECL substructures it contains. A second is the realization problem: determine which combinatorial patterns can be implemented with qubits, qudits, stabilizer codes, or more general isometries. A third direction concerns multi-key QECL, where different overlapping families of qualified sets define different key--clone decompositions inside the same encoding. 

Overall, the conclusion is that quantum encrypted cloning should not be viewed only as a specific protocol. It is a structural primitive arising whenever quantum information is distributed so that several non-qualified subsystems can alternatively complete a common key into qualified recovery sets. This turns quantum secret sharing, and indirectly quantum error correction, into systematic design tools for encrypted quantum redundancy.

%%%%%%%%%%%%%%%%%%%%%%%%%%%%%%%%%%

%\newpage
%+Bibliography

\bibliographystyle{unsrt}
\bibliography{refs}

@article{yamaguchi2026encrypted,
  title={Encrypted Qubits Can Be Cloned},
  author={Yamaguchi, Koji and Kempf, Achim},
  journal={Physical Review Letters},
  volume={136},
  number={1},
  pages={010801},
  year={2026},
  publisher={APS}
}

@article{ceara2026cloningencryptedquantumstates,
      title={Cloning Encrypted Quantum States in Arbitrary Dimensions}, 
      author={Filip-Ioan Ceara},
      journal={arXiv preprint arXiv:2604.04888},
      year={2026}
}

@article{gianini2026encrypted,
      title={Encrypted clones can leak: Classification of informative subsets in Quantum Encrypted Cloning}, 
      author={Gabriele Gianini and Omar Hasan and Corrado Mio and Stelvio Cimato and Ernesto Damiani},
      journal={arXiv preprint arXiv:2604.10155},      
      year={2026},
      eprint={2604.10155},
      archivePrefix={arXiv},
      primaryClass={quant-ph},
      url={https://arxiv.org/abs/2604.10155}
}

@article{gianini2026full,
      title={Full characterization of informative subsets in Quantum Encrypted Cloning}, 
      author={Gabriele Gianini and Stelvio Cimato and Jianyi Lin and Omar Hasan and Ernesto Damiani},
      journal={arXiv preprint arXiv:2605.27421},    
      year={2026},
      eprint={2605.27421},
      archivePrefix={arXiv},
      primaryClass={quant-ph},
      url={https://arxiv.org/abs/2605.27421}
}

@article{bai2026classification,
  title={Classification of informative subsets in quantum encrypted cloning on qudits},
  author={Bai, Chen-Ming and Zhou, Xin-Liang and Luo, Yu},
  journal={arXiv preprint arXiv:2605.11642},
  year={2026}
}

@article{lim2026encrypted,
  title={Encrypted Cloning, Absolute Maximal Entanglement and Quantum Secret Sharing},
  author={Lim, Zheng Liang and Lo, Hoi-Kwong},
  journal={arXiv preprint arXiv:2605.26866},
  year={2026}
}

@article{tassa2007hierarchical,
  author  = {Tassa, Tamir},
  title   = {Hierarchical Threshold Secret Sharing},
  journal = {Journal of Cryptology},
  volume  = {20},
  number  = {2},
  pages   = {237--264},
  year    = {2007},
  doi     = {10.1007/s00145-006-0334-8}
}

@article{brickell1989some,
  author  = {Brickell, Ernest F.},
  title   = {Some Ideal Secret Sharing Schemes},
  journal = {Journal of Combinatorial Mathematics and Combinatorial Computing},
  volume  = {6},
  pages   = {105--113},
  year    = {1989}
}

@article{chen2024ideal,
  author  = {Chen, Qi and Liu, Wen-Ai and Zhang, Liang and Wang, Yue},
  title   = {Ideal Uniform Multipartite Secret Sharing Schemes},
  journal = {Information Sciences},
  volume  = {657},
  pages   = {119958},
  year    = {2024},
  doi     = {10.1016/j.ins.2023.119958}
}

@article{hillery1999quantum,
  author  = {Hillery, Mark and Bu{\v{z}}ek, Vladim{\'i}r and Berthiaume, Andr{\'e}},
  title   = {Quantum Secret Sharing},
  journal = {Physical Review A},
  volume  = {59},
  number  = {3},
  pages   = {1829--1834},
  year    = {1999},
  doi     = {10.1103/PhysRevA.59.1829},
  eprint  = {quant-ph/9806063},
  archivePrefix = {arXiv}
}

@article{cleve1999share,
  author  = {Cleve, Richard and Gottesman, Daniel and Lo, Hoi-Kwong},
  title   = {How to Share a Quantum Secret},
  journal = {Physical Review Letters},
  volume  = {83},
  number  = {3},
  pages   = {648--651},
  year    = {1999},
  doi     = {10.1103/PhysRevLett.83.648},
  eprint  = {quant-ph/9901025},
  archivePrefix = {arXiv}
}

@article{gottesman2000theory,
  author  = {Gottesman, Daniel},
  title   = {Theory of Quantum Secret Sharing},
  journal = {Physical Review A},
  volume  = {61},
  number  = {4},
  pages   = {042311},
  year    = {2000},
  doi     = {10.1103/PhysRevA.61.042311},
  eprint  = {quant-ph/9910067},
  archivePrefix = {arXiv}
}

@article{ogawa2005quantum,
  author  = {Ogawa, Tomohiro and Sasaki, Akira and Iwamoto, Mitsugu and Yamamoto, Hirosuke},
  title   = {Quantum Secret Sharing Schemes and Reversibility of Quantum Operations},
  journal = {Physical Review A},
  volume  = {72},
  number  = {3},
  pages   = {032318},
  year    = {2005},
  doi     = {10.1103/PhysRevA.72.032318},
  eprint  = {quant-ph/0505001},
  archivePrefix = {arXiv}
}

@article{zhang2015strongly,
  author  = {Zhang, Paul and Matsumoto, Ryutaroh},
  title   = {Quantum Strongly Secure Ramp Secret Sharing},
  journal = {Quantum Information Processing},
  volume  = {14},
  pages   = {715--729},
  year    = {2015},
  doi     = {10.1007/s11128-014-0880-7},
  eprint  = {1404.5749},
  archivePrefix = {arXiv}
}

@misc{matsumoto2014coding,
  author        = {Matsumoto, Ryutaroh},
  title         = {Coding Theoretic Construction of Quantum Ramp Secret Sharing},
  year          = {2014},
  eprint        = {1405.0149},
  archivePrefix = {arXiv},
  primaryClass  = {quant-ph}
}

@article{matsumoto2017unitary,
  author  = {Matsumoto, Ryutaroh},
  title   = {Unitary Reconstruction of Secret for Stabilizer-Based Quantum Secret Sharing},
  journal = {Quantum Information Processing},
  volume  = {16},
  pages   = {202},
  year    = {2017},
  doi     = {10.1007/s11128-017-1656-1}
}

@article{matsumoto2020classical,
  author  = {Matsumoto, Ryutaroh},
  title   = {Classical Access Structures of Ramp Secret Sharing Based on Quantum Stabilizer Codes},
  journal = {Quantum Information Processing},
  volume  = {19},
  pages   = {9},
  year    = {2020},
  doi     = {10.1007/s11128-019-2503-3},
  eprint  = {1811.05217},
  archivePrefix = {arXiv}
}

@article{choi2013entanglement,
  author  = {Choi, Ran Hee and Miller, Daniel and Sanders, Barry C.},
  title   = {Entanglement Sharing Protocol via Quantum Error-Correcting Codes},
  journal = {Physical Review A},
  volume  = {87},
  number  = {3},
  pages   = {032319},
  year    = {2013},
  doi     = {10.1103/PhysRevA.87.032319},
  eprint  = {1212.4217},
  archivePrefix = {arXiv}
}

@article{lai2022dynamic,
  author  = {Lai, Hong and Pieprzyk, Josef and Pan, Lei},
  title   = {Dynamic Hierarchical Quantum Secret Sharing Based on the Multiscale Entanglement Renormalization Ansatz},
  journal = {Physical Review A},
  volume  = {106},
  number  = {5},
  pages   = {052403},
  year    = {2022},
  doi     = {10.1103/PhysRevA.106.052403}
}

@article{knill1997theory,
  author  = {Knill, Emanuel and Laflamme, Raymond},
  title   = {Theory of Quantum Error-Correcting Codes},
  journal = {Physical Review A},
  volume  = {55},
  number  = {2},
  pages   = {900--911},
  year    = {1997},
  doi     = {10.1103/PhysRevA.55.900},
  eprint  = {quant-ph/9604034},
  archivePrefix = {arXiv}
}

@book{nielsen2010quantum,
  author    = {Nielsen, Michael A. and Chuang, Isaac L.},
  title     = {Quantum Computation and Quantum Information},
  publisher = {Cambridge University Press},
  edition   = {10th Anniversary Edition},
  year      = {2010},
  doi       = {10.1017/CBO9780511976667}
}

\end{document}